# Nonlocal Potentials and Crystalline Order in One and Two Dimensions

Moorad Alexanian

*Department of Physics and Physical Oceanography*
*University of North Carolina Wilmington, Wilmington, NC 28403-5606*

E-mail: alexanian@uncw.edu

**Abstract.** We revisit the seminal 1968 proof of the absence of crystalline order in two dimensions and analyze the importance played in the quantum theorem by the assumption of local pair potentials. We relax the assumption of local potentials and consider instead nonlocal pair potentials. We show that the $1/k^2$-singularity that occurs in the Bogoliubov inequality, which leads to no crystalline order in two dimensions for local potentials and nonzero temperatures, does not occur for nonlocal potentials. Accordingly, crystalline order in one and two dimensions cannot be ruled out for nonlocal pair potentials at finite temperatures.

**Keywords:** crystalline order, Bogoliubov inequality, quantum theorem, local and nonlocal pair potentials

## 1. Introduction

In 1968, Mermin [1] published a seminal paper on the nonexistence of crystalline order in two dimensions (and the obvious analogous argument in one dimension) for both classical and quantum mechanical particles, albeit, the quantum mechanical case was relegated to a short appendix. The proof relies on the local nature of the pair potential $\Phi(\mathbf{r})$ provided $\Phi(\mathbf{r}) - \lambda r^2 / \nabla^2 \Phi(\mathbf{r})/$ provided is integrable as $r = \infty$ and positive and nonintegrable at $r = 0$, both for $\lambda = 0$ and for some positive $\lambda$. The quantum-mechanical proof of the analogous classical theorem is based on the Bogoliubov inequality. In what follows, we reproduce the quantum mechanical proof and indicate where the assumption regarding the pair potential come in and how the proof is modified for nonlocal pair potentials thus allowing for the existence of crystalline order in one and two dimensions.

In the present paper, we consider a detailed proof of the quantum mechanical case and explicitly show the crucial role the local nature of the pair potential plays in establishing the nonexistence of crystalline order in two dimensions. In Sec. 2, we introduce the quantities that play a role in the proof and the nature of the two-dimensional crystal especially the characterization of the existence of crystalline order. In Sec. 3, we review the two-particle Schrödinger equation for both local, as well as, nonlocal potentials. In Sec. 4, we introduce the Bogoliubov inequality and define the corresponding operators that appear in the inequality and express them in terms of the most general pair potentials, which includes nonlocal potentials. In Sec. 5, we show how any nonlocal potential invalidates the quantum proof of the nonexistence of crystalline order in two dimensions. Finally, in Sec. 6, we summarize our results and discuss the connection of the present work to the existence of Bose-Einstein condensation in one and two dimensions.

## 2. Two-dimensional crystal

Consider a two-dimensional crystal with Bravais lattice generated by $\mathbf{a}_1$ and $\mathbf{a}_2$ in a box spanned by lattice $N_1\mathbf{a}_1$, $N_2\mathbf{a}_2$, viz., the space $\mathbf{r} = x_1 N_1 \mathbf{a}_1 + x_2 N_2 \mathbf{a}_2$, with $0 \leq x_1, x_2 \leq 1$, filled with $N = nN_1N_2$ particles, where $n$ is the number of particles per unit cell. Define the **k**th Fourier component of the density by

$$\rho_{\mathbf{k}} = N^{-1} \langle \hat{\rho}_{\mathbf{k}} \rangle, \tag{1}$$

where

$$\hat{\rho}_{\mathbf{k}} = \int d\mathbf{r} e^{-i\mathbf{k}\cdot\mathbf{r}} \hat{\rho}(\mathbf{r}) = \sum_{i=1}^{N} e^{-i\mathbf{k}\cdot\hat{\mathbf{r}}_i}, \quad (2)$$

since

$$\hat{\rho}(\mathbf{r}) = \sum_{i=1}^{N} \delta(\mathbf{r} - \hat{\mathbf{r}}_i), \quad (3)$$

$$\langle f \rangle = \int d\mathbf{r}_1 \cdots d\mathbf{r}_N e^{-\beta U(\mathbf{r}_1 \cdots \mathbf{r}_N)} f(\mathbf{r}_1 \cdots \mathbf{r}_N)/Q, \quad (4)$$

$$Q = \int d\mathbf{r}_1 \cdots d\mathbf{r}_N e^{-\beta U(\mathbf{r}_1 \cdots \mathbf{r}_N)}, \quad (5)$$

$$U(\mathbf{r}_1 \cdots \mathbf{r}_N) = \frac{1}{2} \sum_{i \neq j} \Phi(\mathbf{r}_1, \mathbf{r}_2), \quad (6)$$

and
where $\beta = 1/k_B T$. The integrations are over all the interior of the box. The pair potential $\Phi(\mathbf{r}_1, \mathbf{r}_2)$ allows for nonlocal potentials. In the case of local potentials, one has that $\Phi(\mathbf{r}_1, \mathbf{r}_2) = \Phi(\mathbf{r}_1 - \mathbf{r}_2)$ (this local potential does not satisfy rotation invariance). If $\mathbf{k}$ is a vector of the lattice reciprocal to that generated by $\mathbf{a}_1$, $\mathbf{a}_2$, that is, $e^{i\mathbf{k}\cdot\mathbf{t}_N} = 1$ for all $N$ with $\mathbf{t}_N = N_1\mathbf{a}_1 + N_2\mathbf{a}_2$, then $\rho_\mathbf{k}$ will in general be nonzero, whereas it will vanish in the thermodynamic limit if $\mathbf{k}$ is not a reciprocal-lattice vector. Accordingly, the criterion for crystallization is

$$\lim \rho_\mathbf{k} = 0, \ \mathbf{k} \text{ not a reciprocal-lattice vector}, \quad (7)$$
$$\lim \rho_\mathbf{k} \neq 0, \text{ for at least one nonzero reciprocal-lattice vector } \mathbf{k},$$

where lim denotes the thermodynamic limit $N_1, N_2 \to \infty$.

## 3. Two-particle Schrödinger equation

The time-dependent Schrödinger equation for two particles and a general two-body potential is

$$i\hbar \frac{\partial \Psi(\mathbf{r}_1, \mathbf{r}_2)}{\partial t} = -\frac{\hbar^2}{2m}\nabla_1^2 \Psi(\mathbf{r}_1, \mathbf{r}_2) - \frac{\hbar^2}{2m}\nabla_2^2 \Psi(\mathbf{r}_1, \mathbf{r}_2) + \int V(\mathbf{r}_1, \mathbf{r}_2, \mathbf{r}_3, \mathbf{r}_4) \Psi(\mathbf{r}_3, \mathbf{r}_4) d\mathbf{r}_3 d\mathbf{r}_4. \quad (8)$$

The two-particle interaction potential $V(\mathbf{r}_1, \mathbf{r}_2, \mathbf{r}_3, \mathbf{r}_4)$ must satisfy the following general conditions: (i) translation invariance, (ii) Galilean invariance, (iii) identical particles, (iv) rotation invariance, (v) space-reflection invariance, (vi) time-reversal invariance, and (vii) Hermiticity and so [2]

$$V(\mathbf{r}_1, \mathbf{r}_2, \mathbf{r}_3, \mathbf{r}_4) = \delta(\mathbf{r}_1 + \mathbf{r}_2 - \mathbf{r}_3 - \mathbf{r}_4)\langle\mathbf{r}_1 - \mathbf{r}_2 / V / \mathbf{r}_3 - \mathbf{r}_4\rangle \quad \text{(nonlocal)}. \quad (9)$$

The potential (9) represents the most general two-particle potential consistent with the above symmetry requirements.

A mathematically simpler potential can be deduced from (9) when

$$\langle \mathbf{r}_1 - \mathbf{r}_2 / V / \mathbf{r}_3 - \mathbf{r}_4 \rangle = \delta(\mathbf{r}_1 - \mathbf{r}_2 - \mathbf{r}_3 + \mathbf{r}_4)\Phi(/\mathbf{r}_1 - \mathbf{r}_2/) \quad (10)$$

and so

$$V(\mathbf{r}_1, \mathbf{r}_2, \mathbf{r}_3, \mathbf{r}_4) = \frac{1}{2}\delta(\mathbf{r}_1 - \mathbf{r}_3)\delta(\mathbf{r}_2 - \mathbf{r}_4)\Phi(/\mathbf{r}_1 - \mathbf{r}_2/) \quad \text{(local)}. \quad (11)$$

Such potentials as (11) are commonly known as local potentials, whereas the more general potential (9) is commonly known as a nonlocal potential.

The solution $\Psi(\mathbf{r}_1, \mathbf{r}_2)$ of the Schrödinger equation (8), for a particular nonlocal potential $V(\mathbf{r}_1, \mathbf{r}_2, \mathbf{r}_3, \mathbf{r}_4)$, allows us to define another nonlocal potential $\Phi(\mathbf{r}_1, \mathbf{r}_2)$ via

$$\Phi(\mathbf{r}_1, \mathbf{r}_2)\Psi(\mathbf{r}_1, \mathbf{r}_2) \equiv \int \langle \mathbf{r}_1 - \mathbf{r}_2 / V / 2\mathbf{r}_3 - \mathbf{r}_1 - \mathbf{r}_2 \rangle \Psi(\mathbf{r}_3, \mathbf{r}_1 + \mathbf{r}_2 - \mathbf{r}_3)d\mathbf{r}_3, \quad (12)$$

with the aid of (9). Accordingly, we introduce nonlocal potentials into the Schrödinger equation via $\Phi(\mathbf{r}_1, \mathbf{r}_2)$ rather than $V(\mathbf{r}_1, \mathbf{r}_2, \mathbf{r}_3, \mathbf{r}_4)$. Note that if, in addition,

$$\langle \mathbf{r}_1 - \mathbf{r}_2 / V / 2\mathbf{r}_3 - \mathbf{r}_1 - \mathbf{r}_2 \rangle = \frac{1}{2}\delta(\mathbf{r}_1 - \mathbf{r}_3)\Phi(/\mathbf{r}_1 - \mathbf{r}_2/), \quad (13)$$

then the integral in (12) gives $\Phi(\mathbf{r}_1, \mathbf{r}_2) = \frac{1}{2}\Phi(/\mathbf{r}_1 - \mathbf{r}_2/)$, which is a local potential.

## 4. Quantum theorem

Consider the Bogoliubov inequality [3, 4]

$$\frac{1}{2}\langle \hat{A}\hat{A}^\dagger + \hat{A}^\dagger \hat{A}\rangle \geq k_B T |\langle [\hat{C}, \hat{A}]\rangle|^2 / \langle [[\hat{C}, \hat{H}], \hat{C}^\dagger]\rangle, \quad (14)$$

where the braces indicate full canonical ensemble averages.
Let

$$\hat{A} = \sum_i e^{i(\mathbf{k}+\mathbf{K})\cdot \hat{\mathbf{r}}_i} = \hat{\rho}_{\mathbf{k}+\mathbf{K}}, \quad (15)$$

and

$$\hat{C} = \frac{1}{2}\sum_i [\hat{\mathbf{p}}_i \sin(\mathbf{k}\cdot\hat{\mathbf{r}}_i) + \sin(\mathbf{k}\cdot\hat{\mathbf{r}}_i)\hat{\mathbf{p}}_i]. \quad (16)$$

We obtain that

$$\langle \hat{A}\hat{A}^\dagger + \hat{A}^\dagger \hat{A}\rangle = 2\langle \hat{\rho}_{\mathbf{k}+\mathbf{K}}\,\hat{\rho}_{-\mathbf{k}-\mathbf{K}}\rangle, \quad (17)$$

and

$$|\langle [\hat{C}, \hat{A}]\rangle|^2 = \frac{\hbar^2}{4}(\mathbf{k}+\mathbf{K})^2 |\langle \hat{\rho}_{2\mathbf{k}+\mathbf{K}} - \hat{\rho}_{\mathbf{K}}\rangle|^2. \quad (18)$$

The Hamiltonian of the two-particle system is

$$\hat{H} = \frac{1}{2m}\sum_i \hat{p}_i^2 + \hat{U}(\hat{\mathbf{r}}_i \cdots \hat{\mathbf{r}}_N), \quad (19)$$

where

$$\hat{U}(\hat{\mathbf{r}}_i \cdots \hat{\mathbf{r}}_N) = \frac{1}{2}\sum_{i\neq j}\Phi(\hat{\mathbf{r}}_i, \hat{\mathbf{r}}_j). \quad (20)$$

One obtains, for the kinetic energy term of the denominator of (14)

where the ensemble average of the total momentum is set equal to zero, i.e., $\langle \sum_i \hat{p}_i\rangle = 0$. Result

$$\langle [[\hat{C}, \frac{1}{2m}\sum_i \hat{p}_i^2], \hat{C}^\dagger]\rangle = \frac{1}{m}\sum_i \langle \hbar^2(\mathbf{k}\cdot\hat{p}_i)^2 + \hbar^2 k^2 \hat{p}_i \cdot \cos^2(\mathbf{k}\cdot\hat{\mathbf{r}}_i)\hat{p}_i + \frac{1}{4}\hbar^4 k^4 \sin^2(\mathbf{k}\cdot\hat{\mathbf{r}}_i)\rangle \quad (21)$$

(21) agrees with that given by Mermin [1].

We next obtain the term in the denominator associated with the potential without making the assumption made by Mermin of a local potential.

$$\langle[[\hat{C},\hat{U}],\hat{C}^{\dagger}]\rangle = \frac{\hbar^2}{2}\sum_{i\neq j}\langle\Big[\sin(\mathbf{k}\cdot\mathbf{r}_i)\cos(\mathbf{k}\cdot\mathbf{r}_i)\mathbf{k}\cdot\nabla_i + \sin(\mathbf{k}\cdot\mathbf{r}_j)\cos(\mathbf{k}\cdot\mathbf{r}_j)\mathbf{k}\cdot\nabla_j$$
$$+ \sin^2(\mathbf{k}\cdot\mathbf{r}_i)\nabla_i\cdot\nabla_i + \sin^2(\mathbf{k}\cdot\mathbf{r}_j)\nabla_j\cdot\nabla_j + 2\sin(\mathbf{k}\cdot\mathbf{r}_i)\sin(\mathbf{k}\cdot\mathbf{r}_j)\nabla_i\cdot\nabla_j\Big]\Phi(\mathbf{r}_i,\mathbf{r}_j)\rangle. \quad (22)$$

## 5. Crystalline order

The proof of the absence of crystalline order in two dimensions [1] at nonzero temperatures is based on the interparticle potential being local, viz., $\Phi(\mathbf{r}_i, \mathbf{r}_j) = \Phi(\mathbf{r}_i - \mathbf{r}_j)$. Accordingly, $\nabla_j\Phi(\mathbf{r}_i - \mathbf{r}_j) = -\nabla_i\Phi(\mathbf{r}_i - \mathbf{r}_j)$ and so, (22) becomes

$$\langle[[\hat{C},\hat{U}],\hat{C}^{\dagger}]\rangle = \frac{\hbar^2}{2}\sum_{i\neq j}\langle[\frac{1}{2}(\sin 2\mathbf{k}\cdot\mathbf{r}_i - \sin 2\mathbf{k}\cdot\mathbf{r}_j)\mathbf{k}\cdot\nabla_i + (\sin\mathbf{k}\cdot\mathbf{r}_i - \sin\mathbf{k}\cdot\mathbf{r}_j)^2\nabla_i\cdot\nabla_i]\Phi(\mathbf{r}_i - \mathbf{r}_j)\rangle, \quad (23)$$

which agrees with Mermin's Eq.(39) for the potential term [1].

The kinetic term (21) is bounded from above by

$$|\langle[[\hat{C},\frac{1}{2m}\sum_i\hat{p}_i^2],\hat{C}^{\dagger}]\rangle| \leq \hbar^2 k^2 N(4t + \frac{1}{4m}\hbar^2 k^2), \quad (24)$$

where $t$ is the kinetic energy per particle in equilibrium while the potential energy term (23) is bounded from above by

$$|\langle[[\hat{C},\hat{U}],\hat{C}^{\dagger}]\rangle| \leq \hbar^2 k^2 \sum_{i\neq j}[\frac{1}{4}|\mathbf{r}_{ij}||\nabla_i\Phi(\mathbf{r}_{ij})| + 2|\mathbf{r}_{ij}|^2|\nabla_i^2\Phi(\mathbf{r}_{ij})|], \quad (25)$$

where $\mathbf{r}_{ij} = \mathbf{r}_i - \mathbf{r}_j$ and we have used the inequality $/\sin(x) \pm \sin(y)| \leq |x \pm y/$. The Bogoliubov inequality (14) becomes

$$\frac{1}{N}\langle\hat{\rho}_{\mathbf{k}+\mathbf{K}}\hat{\rho}_{-\mathbf{k}-\mathbf{K}}\rangle \geq \frac{k_B T(\mathbf{k}+\mathbf{K})^2(\rho_{2\mathbf{k}+\mathbf{K}} - \rho_{\mathbf{K}})^2}{4k^2(4t + \hbar^2 k^2/4m + (1/N)\sum_{i\neq j}[(1/4)|\mathbf{r}_{ij}||\nabla_i\Phi(\mathbf{r}_{ij})| + 2|\mathbf{r}_{ij}|^2|\nabla_i^2\Phi(\mathbf{r}_{ij})|])}. \quad (26)$$

Mermin [1] has shown that all the terms in (26) are bounded in the thermodynamic limit for the appropriate local potential. Crystallization occurs if both conditions in (7) are satisfied. We will assume crystallization, viz. (7), and will arrive at a contradiction, viz., $\rho_{2\mathbf{k}+\mathbf{K}} \to 0$. Note the $1/k^2$-singularity in (26), therefore, on integrating (26) for $k$ in the range 0 to $\epsilon > 0$, we note that the integral on the right-hand-side of (26) diverges logarithmically at $k = 0$ for a two-dimensional system while the integral on the left-hand-side of (26) is finite and so if $\lim \rho_\mathbf{k} = 0$, for $\mathbf{k}$ not a reciprocal-lattice vector, then $\rho_\mathbf{k} = 0$, for at least one non-zero lattice vector $\mathbf{k}$. Accordingly, there is no crystalline order in two dimensions for nonzero temperatures.

It is important to remark how significant is the assumption of a local pair potential in order to arrive at such a quantum theorem of the absence of crystalline order in two dimensions. In the case of nonlocal potentials, the inequality associated with the kinetic energy remains the same as in the case of local potentials. However, for the potential energy (22), the upper bound is given for nonlocal potentials by

$$|\langle[[\hat{C},\hat{U}],\hat{C}^{\dagger}]\rangle| \leq \frac{\hbar^2}{2}\sum_{i\neq j}\Big[|\nabla_i^2\Phi(\mathbf{r}_i,\mathbf{r}_j)| + |\nabla_j^2\Phi(\mathbf{r}_i,\mathbf{r}_j)| + 2|\nabla_{ij}\Phi(\mathbf{r}_i,\mathbf{r}_j)| + k(|\nabla_i\Phi(\mathbf{r}_i,\mathbf{r}_j)| + |\nabla_j\Phi(\mathbf{r}_i,\mathbf{r}_j)|)\Big]. \quad (27)$$

Accordingly, for nonlocal potentials the $1/k^2$-singularity is removed and so crystallization can occur in both one and two dimensions.

## 6. Conclusions

The existence or nonexistence of crystalline order in one and two dimensions for quantum systems for finite temperatures hinges heavily on the nature of the pair potential. The most general pair potentials are the so-called nonlocal potentials. The nonexistence of crystalline order in two dimension is based on the local nature of the assumed pair potential. We show that such nonexistence proof of crystalline order in one and two dimensions fails for nonlocal pair potentials. The removal of the $1/k^2$-singularity in the Bogoliubov inequality for the proof of crystallization is reminiscent of the proof for the existence of Bose-Einstein condensation in one and two dimensions. The removal of the $1/k^2$-singularity is quite analogous to our present case of crystallization and leads in boson fluids to a nonuniform Bose-Einstein condensate [2, 5].

**References**

[1] N.D. Mermin, Phys. Rev. **176** (1968) 250.
[2] M. Alexanian, Phys. Rev. A **4** (1971) 1684.
[3] N.D. Mermin, H. Wagner, Phys. Rev. Lett. **17** (1966) 1133.
[4] H. Wagner, Z. Physik **195** (1966) 273.
[5] M. Alexanian, Arm. J. Phys. **12** (2019) 185.